\documentstyle[amssymb,aps,epsfig,float,twocolumn]{revtex}
\restylefloat{figure}
\title{Quantum to classical crossover in the 2D easy-plane XXZ model}
\author{ H.~Fehske, C. Schindelin, A. Wei{\ss}e, H.~B\"uttner}
\address{Physikalisches Institut, Universit\"at Bayreuth, 
  D-95440 Bayreuth, Germany}
\author{D.~Ihle}
\address{Institut f\"ur Theoretische Physik, Universit\"at Leipzig,
  D-04109 Leipzig, Germany \\{\rm (\today)}
  }
\address{~\parbox{14cm}{\rm
    \medskip
Ground-state and thermodynamical properties of the spin-1/2 
two-dimensional easy-plane XXZ model are investigated
by both a Green's-function approach and by Lanczos 
diagonalizations on lattices with up to 36 sites. 
We calculate the spatial and temperature dependences 
of various spin correlation functions,
as well as the wave-vector dependence of the spin susceptibility
for all anisotropy parameters  $\Delta$. 
In the easy--plane ferromagnetic 
region $(-1< \Delta < 0)$, the longitudinal correlators of spins 
at distance $r$ change sign at a finite temperature $T_0(\Delta, {\bf r})$.
This transition, observed in the 2D case for the first time,
can be interpreted as a quantum to classical crossover.
 }}
\begin{document}
\maketitle
\section{Introduction}
The magnetic properties of low-dimensional quantum spin systems with
spin anisotropy, such as the quasi-one-dimensional (1D) 
cuprates~\cite{Jo97} and the quasi-2D high-$T_c$ parent 
compounds~\cite{MRB99}, are of growing interest.
The $S=1/2$ XXZ model
\begin{equation}
  {\cal H} = \frac{J}{2}
\sum_{\langle i,j\rangle} (S_i^+S_j^- +\Delta  S_i^z S_j^z) 
\label{xxz}
\end{equation} 
($\langle i,j\rangle$ denote nearest-neighbor (NN) sites; throughout
we set $J=1$) usually serves as the generic model for those systems. 
Recently, in the ferromagnetic (FM) region  
($-1<\Delta<0$) of the 1D model a quantum-classical crossover 
in the longitudinal spin correlators 
was found by means of exact diagonalization (ED)~\cite{FM99} 
and a Green's-function theory~\cite{SFBI99}. 
For the XXZ model on a square lattice, an analytical 
approach to the spin susceptibility taking into account 
the magnetic short-range order  (SRO) at arbitrary temperatures 
does not yet exist.

In this contribution the spin correlations in the easy-plane region  
$-1<\Delta<1$ of the 2D XXZ model are examined by both a Green's-function 
theory outlined in the Appendix and by exact finite-cluster diagonalizations 
of the model~(\ref{xxz}) on lattices with up to 36 spins using 
periodic boundary conditions. 
We mainly focus on the characteristics of a possible quantum 
to classical crossover in the FM regime.  
Moreover, for the first time, the complete wave-vector, 
temperature and $\Delta$ dependences of the static 
transverse and longitudinal spin susceptibilities are calculated.
\section{Ground-state properties}
\begin{figure}[!htb] 
\epsfig{file= 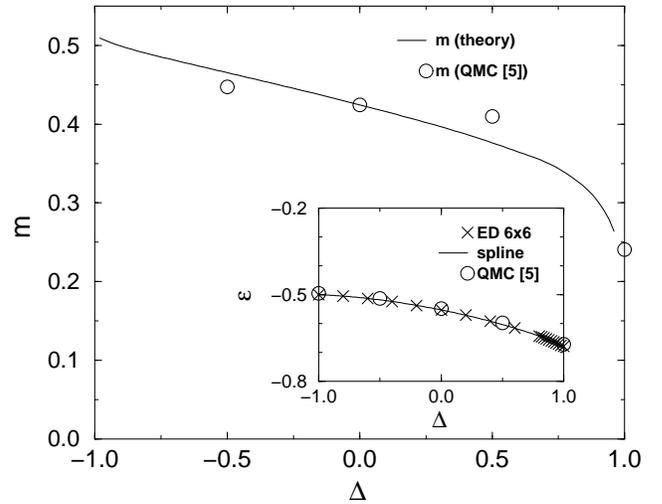, width =.98 \linewidth}
\caption{Magnetization $m$ and ground-state energy $\varepsilon$ of the
2D easy-plane XXZ model.}
\label{fig1}
\end{figure}

In Fig.~\ref{fig1} our results for the magnetization $m(\Delta)$ are 
compared with available quantum Monte Carlo (QMC) data~\cite{OK88}, where
the ED/QMC data for the ground-state energy per site $\varepsilon(\Delta)$
(inset) is taken as input for the Green's-function approach 
($C_{10}^{zz}=\frac{1}{2}\partial\varepsilon/
\partial\Delta$, $C_{10}^{+-}=\varepsilon/2-\Delta C_{10}^{zz}$).

\begin{figure}[!htb] 
\epsfig{file= 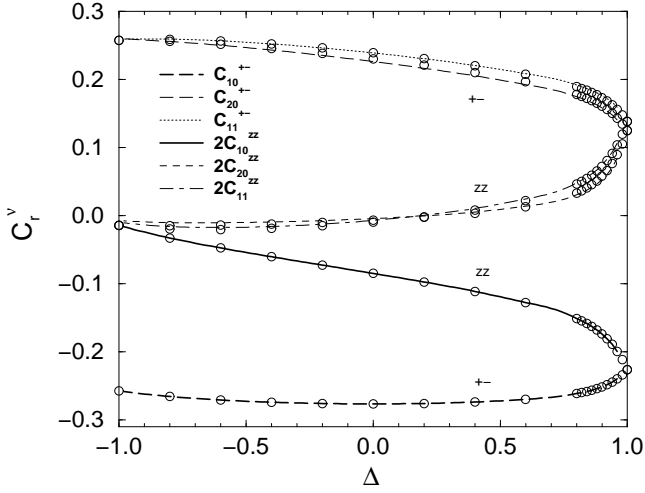, width = .98\linewidth}
\caption{Transverse and longitudinal spin correlation functions
$C_{\bf r}^{\nu}$ at $T=0$. Symbols denote ED results obtained 
for a 6$\times$6 lattice.}
\label{fig2}
\end{figure}

As can be seen from Fig.~\ref{fig2}, the short-ranged correlations 
calculated analytically are in excellent agreement with our ED data. 
Let us stress that the finite-size dependence 
of the ED data is almost negligible by going from a 32- to a 
36-site lattice. At $\Delta=1$ the rotational symmetry 
$C_{\bf r}^{+-}=2C_{\bf r}^{zz}$ is visible. At the quantum critical point
$\Delta =-1$ we have  $C^{+-}_{{\bf r},\tilde{\cal H}}=2C_{\bf r}^{zz} =1/6$ 
(cf. Eq.~(\ref{utrafox2})). 
The non-analytical limiting behavior 
$\lim_{\Delta \to -1^+} C_{\bf r}^{zz}=0$ results from both
the QMC~\cite{OK88} and ED data (obtained in the subspace with total 
spin projection $S^z=0$).

The static spin susceptibilities $\chi_{\bf q}^{\nu}(\Delta)$ are depicted 
in Fig.~3.  In the FM region, for sufficiently low $\Delta$ values,
$\chi_{\bf q}^{zz}$ shows a maximum at ${\bf q}=0$ being a precursor of 
the FM instability (in the $zz$-correlators) at $\Delta=-1$. Note that  
$(\chi_{\bf Q}^{+-})^{-1}=0$, reflecting the transverse 
long-range order (LRO) at $T=0$, by Eq.~(\ref{utrafox1}) 
corresponds to $(\chi_{0,\tilde{\cal H}}^{+-})^{-1}=0$. 
In the antiferromagnetic (AFM) 
region $0<\Delta<1$ the maximum in $\chi_{\bf q}^{zz}$ at ${\bf q}={\bf Q}$
is indicative of the longitudinal AFM LRO at $\Delta\geq 1$.

Finally, in Fig.~\ref{fig4} we show the longitudinal spin-wave spectrum
$\omega_{\bf q}^{zz}$ (cf. Eq.~(\ref{wzz})). For $q\equiv |{\bf q}|\ll 1$
we have  $\omega_{\bf q}^{zz}=c_s^{zz} q$, where the spin-wave velocity
$c_s^{zz}$ increases with $\Delta$ over the whole easy-plane
region. The minimum in $\omega_{\bf q}^{zz}$ at ${\bf q}={\bf Q}$ in 
the AFM region corresponds to the maximum in $\chi_{\bf q}^{zz}$
(cf. Fig.~3 a) and reflects the increase of the longitudinal
AFM SRO with $\Delta$ (see also $C_{\bf r}^{zz}(\Delta)$ in Fig.~\ref{fig2}).
\begin{figure}[!htb] 
\epsfig{file= 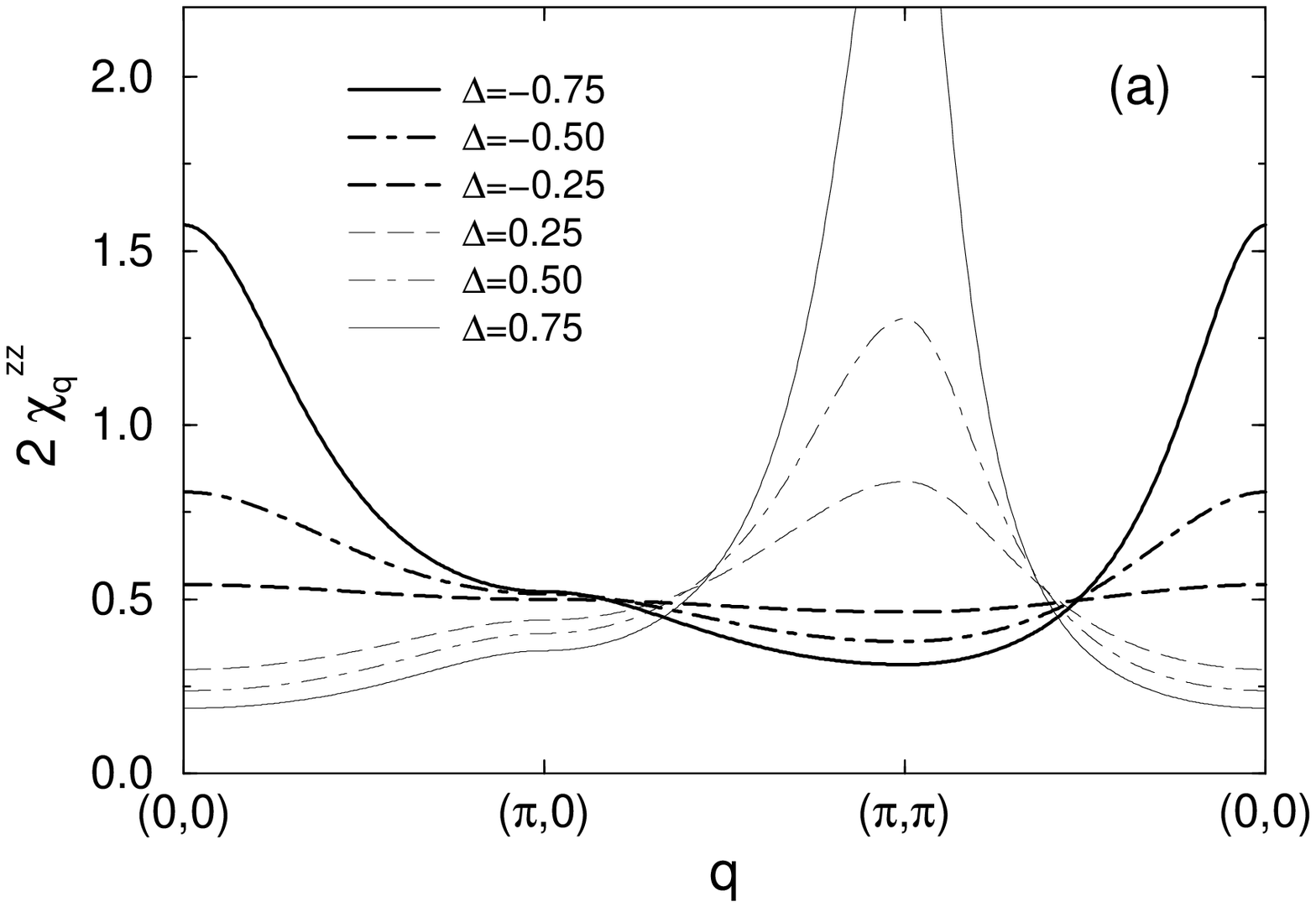, width =.98\linewidth}\\
\epsfig{file= 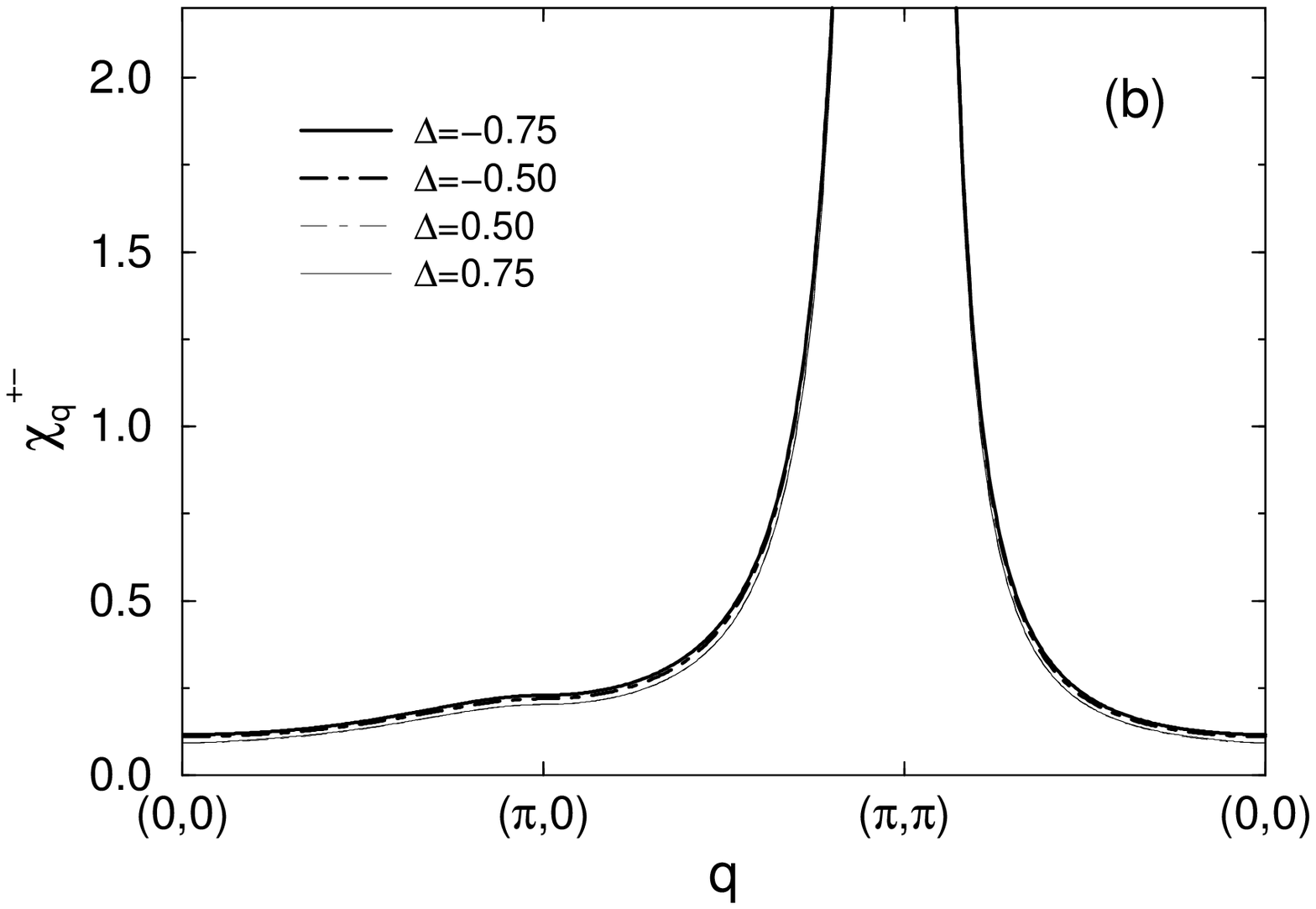, width =.98\linewidth}
\caption{Wave-vector dependence of the longitudinal (a) and
transverse (b) static susceptibilities $\chi_{\bf q}^{\nu}$ at $T=0$.}
\label{fig3}
\end{figure}
\begin{figure}[!htb] 
\epsfig{file= 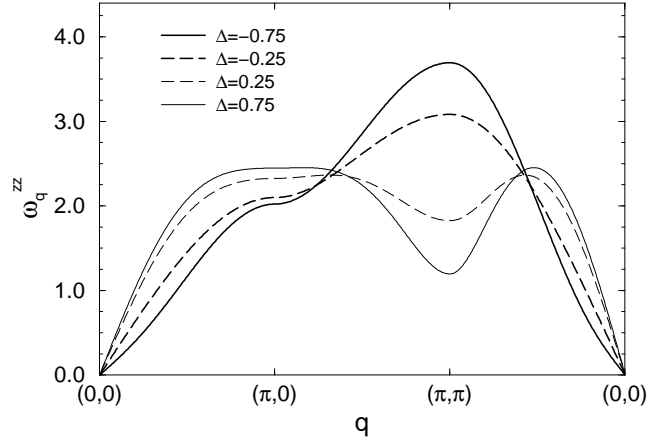, width =.98\linewidth}
\vspace*{-0.2cm}
\caption{Longitudinal spin-wave dispersion $\omega_{\bf q}^{zz}$ along the 
major symmetry directions of the 2D Brillouin zone.}
\label{fig4}
\end{figure}\vspace*{-0.2cm}
\section{Finite-temperature results}
\begin{figure}[!b] 
\epsfig{file= 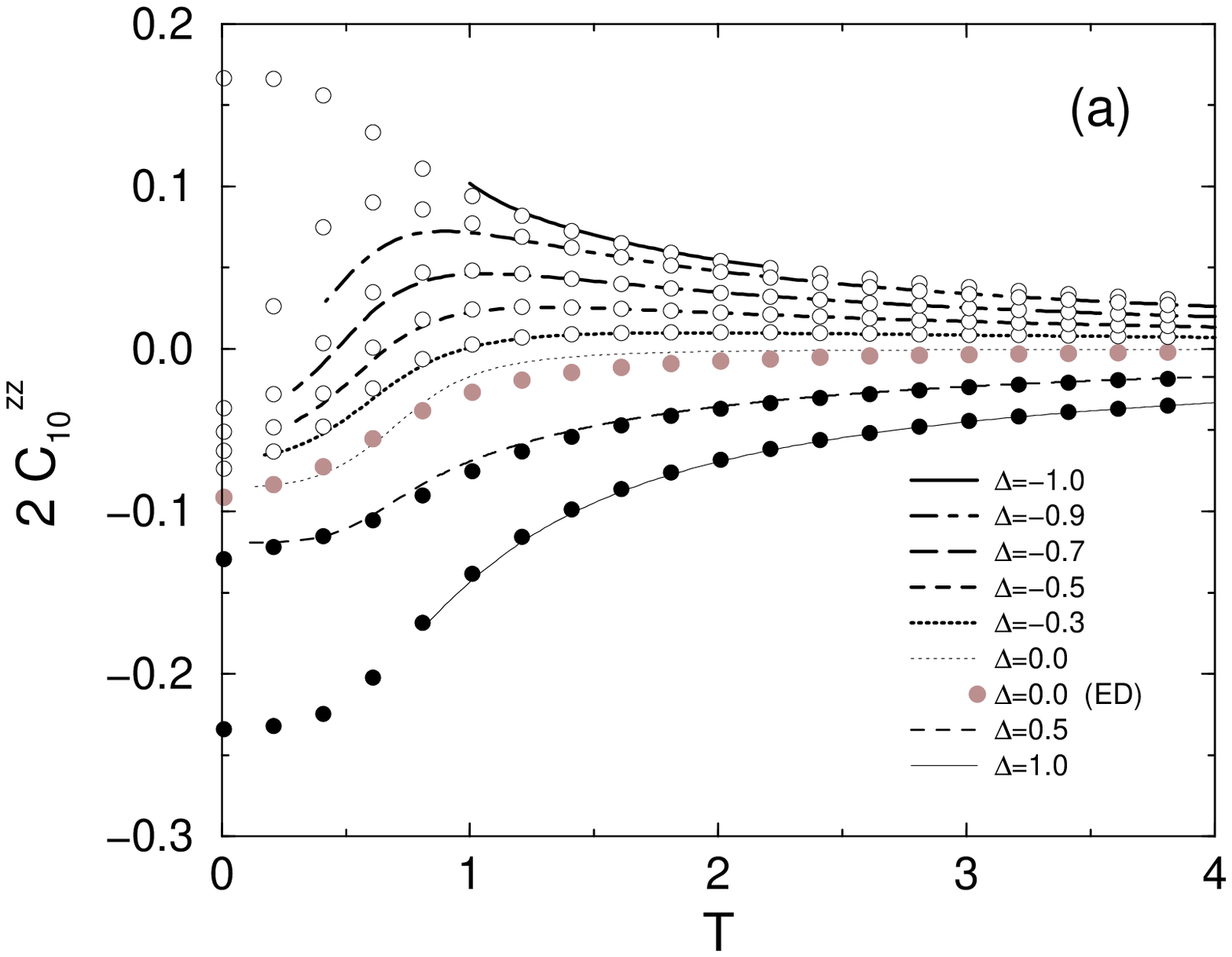, width =.98\linewidth}\\
\epsfig{file= 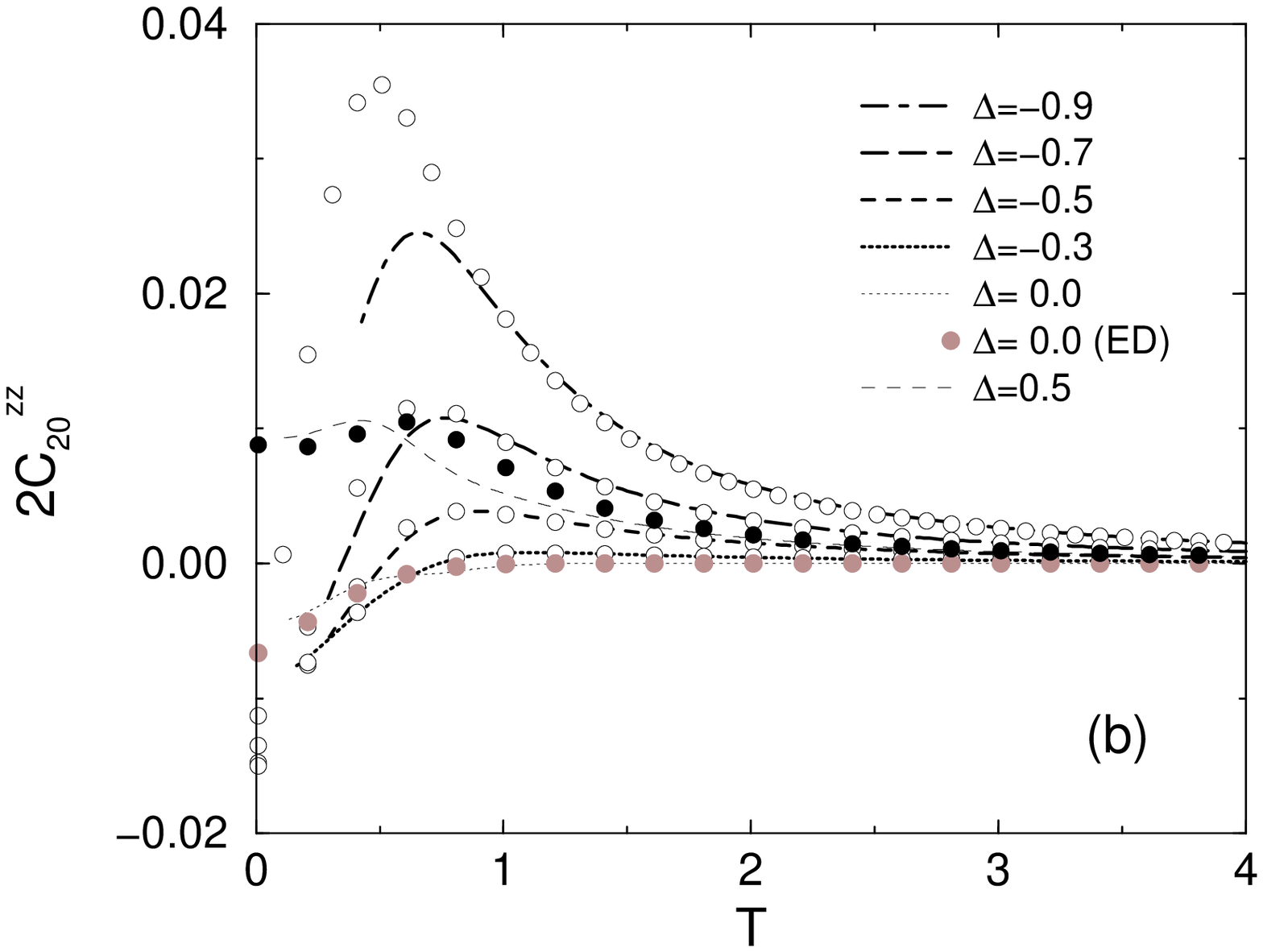, width =.98\linewidth}
\caption{Temperature dependence of the NN (a) and next NN (b) longitudinal
spin correlation functions $C_{\bf r}^{zz}$. 
Symbols denote ED results obtained for a 4$\times$4 lattice.}
\label{fig5}
\end{figure}

The temperature dependence of the short-ranged longitudinal
spin correlations is displayed in Fig.~\ref{fig5}. Again the analytical
results agree remarkably well with the ED data. 
In the FM region, for the first time in the 2D model, 
we observe  the so-called ``sign-changing'' effect 
which was found numerically~\cite{FM99} in the 1D model 
and later on reproduced  by our Green's-function calculations~\cite{SFBI99}. 
That is, at fixed separation $r$ and with increasing temperature
or at fixed temperature and with increasing $r$, $C_{\bf r}^{zz}$
changes sign from negative to positive values. The temperature 
$T_0(\Delta,{\bf r})$ where $C_{\bf r}^{zz}(T_0(\Delta,{\bf r}),\Delta)=0$
are given in Table~I. As in the 1D case, $T_0$ at fixed $\Delta$ decreases
with increasing $r$.  However, compared to the 1D case~\cite{SFBI99},
our analytical results are in much better agreement with the ED data. 
The sign change of $C_{\bf r}^{zz}$ 
may be interpreted as a quantum to classical crossover~\cite{FM99}
because with increasing temperature the system behaves more classically,
i.e., it becomes dominated by the potential energy (negative
$\Delta$ term of the Hamiltonian favoring the parallel alignment of 
two spins). In the AFM region we obtain the expected alternating signs
of  $C_{\bf r}^{zz}$ corresponding to the longitudinal AFM SRO.

\begin{figure}[!b] 
\epsfig{file= 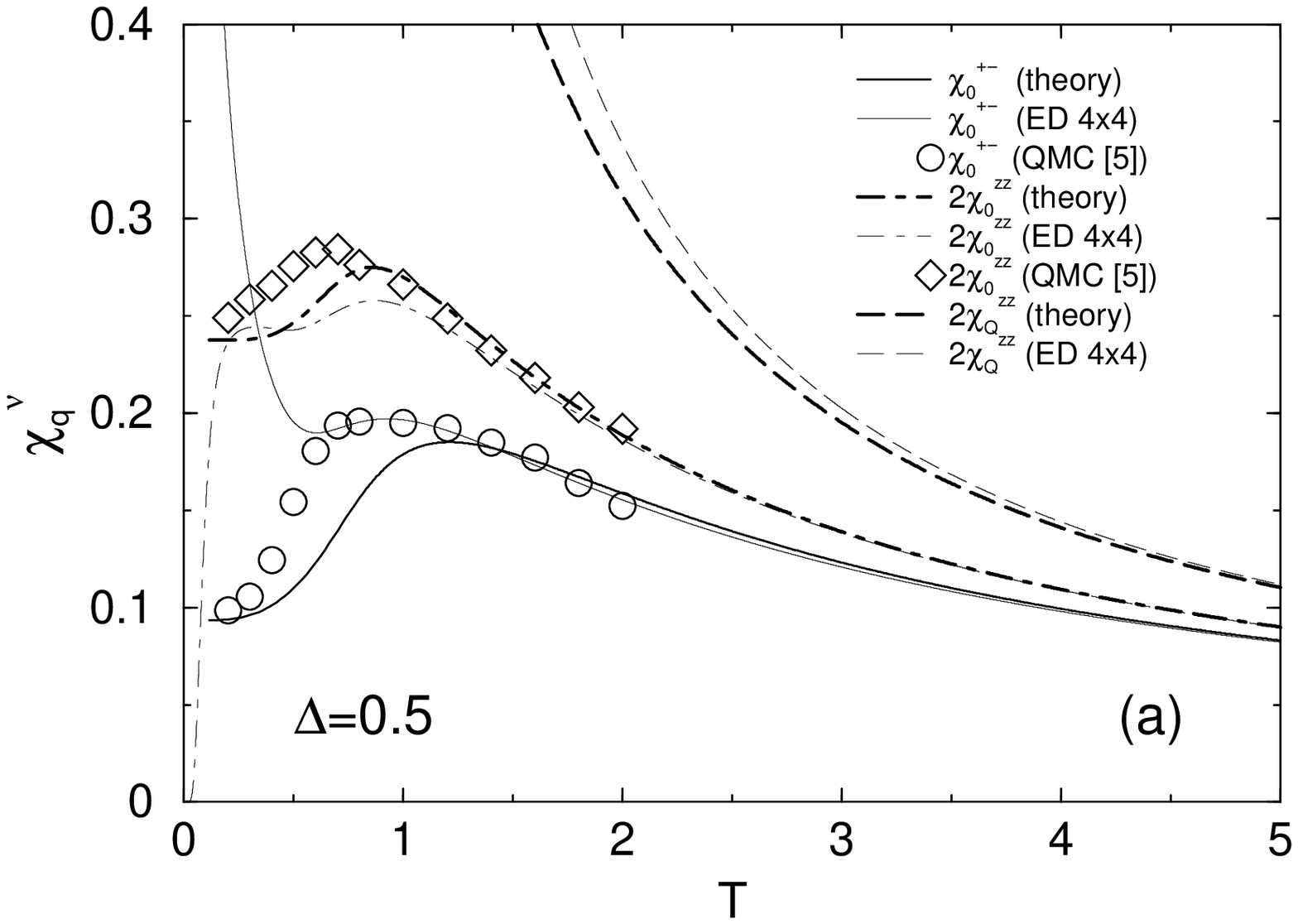, width =.98\linewidth}\\
\epsfig{file= 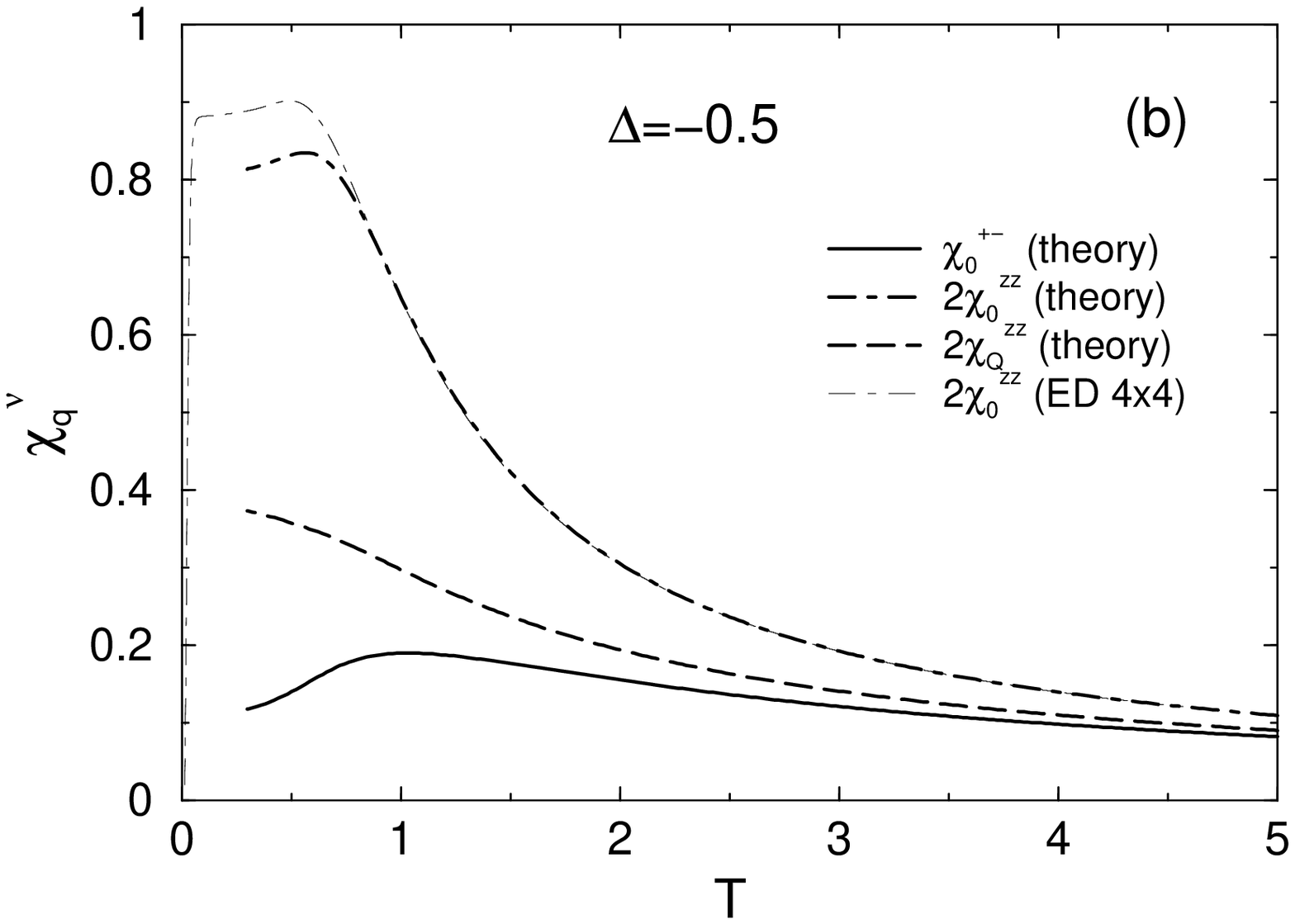, width =.98\linewidth}
\caption{Longitudinal and transverse static spin susceptibilities 
$\chi_{\bf q}^{\nu}$ as functions of temperature $T$ 
for the 2D AFM (a) and FM (b) easy-plane XXZ models.}
\label{fig6}
\end{figure}

In Fig.~\ref{fig6} various susceptibilities 
$\chi_{\bf q}^{\nu}$ at ${\bf q}=0, {\bf Q}$
are plotted as functions of $T$ and compared with numerical data. 
For $\Delta =0.5$ the longitudinal and transverse uniform susceptibilities
are in reasonable agreement with the QMC results~\cite{OK88} and our ED data 
(the up- and downturn at lower temperatures is a finite-size effect).
The increase of $\chi_{0}^{\nu}(T)$, the maximum near the exchange energy
($J=1$), and the crossover to the Curie-Weiss law are due to the decrease 
of AFM SRO with increasing temperature. On the other hand, the staggered
susceptibility $\chi_{\bf Q }^{zz}$ is enhanced as compared with 
$\chi_{0}^{zz}$ by the longitudinal AFM SRO. In the FM region
(Fig.~\ref{fig6}~b, $\Delta=-0.5$) the maximum in  $\chi_{0}^{zz}$, where the
analytical and numerical results yield nearly the same position,
may be explained as a combined SRO and sign changing effect as
discussed for the 1D model in Ref.~\cite{SFBI99}. Contrary to the 
AFM region, $\chi_{\bf Q }^{zz}$ is suppressed  as compared with 
$\chi_{0}^{zz}$ which is caused by the FM correlations above $T_0$.
The temperature dependence of 
$\chi_{0 }^{+-}=\chi_{{\bf Q},\tilde{\cal H}}^{+-}$
may be explained again as a SRO effect. 
Here, the transverse FM SRO results in a spin stiffness against
the orientation of the transverse spin components along a staggered field
perpendicular to the $z$-direction, so that 
$\chi_{{\bf Q},\tilde{\cal H}}^{+-}$ is suppressed at low temperatures
and exhibits a maximum.
\vspace*{-0.2cm}
\begin{table}
\caption{Temperature $T_0(\Delta;{\bf r})$ of the sign change in the 
longitudinal correlation functions $C^{zz}_{\bf r}(T;\Delta)$. 
The corresponding results obtained
from ED of a 4$\times$4 lattice are given in parenthesis.\\}
\begin{tabular}{cccc} 
  \rule{0mm}{4mm}$\Delta$ 
 & \multicolumn{3}{c}{$T_0(\Delta;{\bf r})$}  \\
 \rule{0mm}{4mm} & ${\bf r}=(1,0)$ 
& ${\bf r}=(1,1)$ & ${\bf r}=(2,0)$\\[0.1cm] \hline
  -0.1 & 2.98 [2.540] &$\!\!$ 1.76
  & 1.76 [1.520] \rule{0mm}{4mm} \\[0.1cm] 
 -0.3 & 0.96 [0.931] &  0.74 & 0.72 [0.713] \\[0.1cm] 
 -0.5 & 0.66 [0.605] &  0.52 [0.527]& 0.50 [0.476] \\[0.1cm] 
 -0.7 & 0.46 [0.391] &  0.36 [0.303]& 0.34 [0.301] \\[0.1cm]
 -0.9 &$<$0.2 [0.125] & $<$0.2 [0.106]&$<$0.2 [0.106] 
\end{tabular}
\end{table}
\section{Summary}
To resume, we presented a Green's-function theory of magnetic
LRO and SRO in the 2D easy-plane XXZ model which allows the 
complete calculation of all static magnetic properties in excellent 
agreement with numerical diagonalization data. In particular, in 
the FM region we found a quantum to classical crossover in 
the longitudinal spin correlations. We conclude that our approach 
is promising to be applied to other anisotropic spin models,
such as the quasi-2D XXZ model for the  parent compounds of 
high-$T_c$ superconductors.
\section*{Appendix: Green's-function theory}
The spin susceptibilities 
$\chi^{+-}_{\bf q}(\omega)=
-\langle\langle S_{\bf q}^+;S_{- {\bf q}}^-\rangle\rangle_{\omega}$
and  
$\chi^{zz}_{\bf q}(\omega)=-\langle\langle 
S_{\bf q}^z;S_{- {\bf q}}^z\rangle\rangle_{\omega}$,
expressed in terms of two-time retarded commutator Green's functions, 
are determined by the projection method, developed, for the XXZ chain, 
in Ref.~\cite{SFBI99}. Taking the two-operator basis 
$(S_{\bf q}^+, i \dot{S}_{\bf q}^+)^T$ and $(S_{\bf q}^z, i 
\dot{S}_{\bf q}^z)^T$ we obtain
\begin{equation}
\label{chinu}
\chi^{\nu}_{\bf q}(\omega)=-\frac{M^{\nu}_{\bf q}}{\omega^2
-(\omega_{\bf q}^{\nu})^2}\,;\;\;\nu=+-,\,zz,
\end{equation}
with 
\begin{eqnarray}\label{mq1}
M_{\bf q}^{+-}&=&-4 [C_{10}^{+-}(1-\Delta\gamma_{\bf q})
+2C_{10}^{zz} (\Delta-\gamma_{\bf q} )]\,,\\
M_{\bf q}^{zz}&=&-4 C_{10}^{+-}(1-\gamma_{\bf q} )\,, 
\label{mq2}
\end{eqnarray}
$C_{nm}^{\nu}\equiv C_{\bf r}^{\nu}$, 
$C_{\bf r}^{+-}=\langle S_0^+ S_{\bf r}^- \rangle$,
$C_{\bf r}^{zz}=\langle S_0^z S_{\bf r}^z \rangle$,
${\bf r}= n {\bf e}_x +m  {\bf e}_y$, and
$\gamma_{\bf q}=(\cos q_x +\cos q_y)/2$. 
The spin correlators are obtained from Eq.~(\ref{chinu}) as  
\begin{equation}
\label{cn}
C_{\bf r}^{\nu}=\frac{1}{N}\sum_{\bf q}\frac{M_{\bf q}^{\nu}}{2 
\omega_{\bf q}^{\nu}}
 [1+2p(\omega_{\bf q}^{\nu})] \mbox{e}^{i{\bf q r}}\,,
\end{equation}
where $p(\omega_{\bf q}^{\nu})=(\mbox{e}^{\omega_{\bf q}^{\nu}/T}-1)^{-1}$. 
The spectra $\omega_{\bf q}^{\nu}$, calculated in the approximations 
$-\ddot{S}^+_{\bf q}=(\omega_{\bf q}^{+-})^2 S^+_{\bf q}$ and 
$-\ddot{S}^{z}_{\bf q}=(\omega_{\bf q}^{zz})^2 S^z_{\bf q}$ 
introducing vertex parameters $\alpha_i^{\nu}$ ($i=1,2$),
are given by 
\begin{eqnarray}
\label{w+-}
(\omega_{\bf q}^{+-})^2&=&[(1+2\alpha_2^{+-}( C_{20}^{+-} +2 C_{11}^{+-} )]
(1-\Delta  \gamma_{\bf q} )\nonumber\\[0.1cm]
&&\;\;+\Delta (1+4\alpha_2^{+-}(C_{20}^{zz} +2 C_{11}^{zz} )]
(\Delta - \gamma_{\bf q} )\nonumber\\[0.1cm]
&&\;\;+2\alpha_1^{+-}[C_{10}^{+-}(4 \Delta  \gamma_{\bf q}^2 -\Delta
- 3 \gamma_{\bf q})\nonumber\\[0.1cm]
&&\;\;+2 C_{10}^{zz}(4 \gamma_{\bf q}^2 -1 -3 \Delta 
\gamma_{\bf q} )]\,,\\[0.2cm]
\label{wzz}
(\omega_{\bf q}^{zz})^2&=&2(1-\gamma_{\bf q} )[1+2\alpha_2^{zz}
(C_{20}^{+-}+2 C_{11}^{+-})\nonumber\\[0.1cm]
&&\;\;-2\Delta\alpha_1^{zz}C_{10}^{+-}(1 + 4\gamma_{\bf q} )]\,.
\end{eqnarray}
In the easy-plane region $-1<\Delta<1$, the long-range order at 
$T=0$ is reflected in our theory by $\omega_{\bf Q}^{+-}=0$
[${\bf Q}=(\pi,\pi)$]. Accordingly, the condensation part
$C\mbox{e}^{i{\bf Q r}}$ is separated from $C_{\bf r}^{+-}$
(cf. Eq.~(\ref{cn}), and the magnetization $m$ is calculated
as
\begin{equation}
\label{m}
m^2=\frac{1}{N}\sum_{\bf r}C_{\bf r}^{+-} \mbox{e}^{-i{\bf Q r}} =C\,.
\end{equation}
The parameters $\alpha_1^{\nu}(T)$ are determined from the sum rules
$C_{00}^{+-}=1/2$ and $C_{00}^{zz}=1/4$. To obtain $\alpha_2^{\nu}(T)$
we adjust  $C_{10}^{\nu}(T=0)$ taken from our ED data and assume,
as additional conditions  for the calculation 
of $\chi_{\bf q}^{zz}(\omega)$ and $\chi^{+-}_{\bf q}(\omega)$,
temperature independent ratios 
\begin{equation}
\label{rnb}
R^{zz}=\frac{\alpha_2^{zz}(T) -1}{\alpha_1^{zz}(T) -1}
\end{equation} 
and 
\begin{eqnarray}
\label{ratios}
R_{>}^{+-}&=&\frac{\alpha_2^{+-}(T)-1}{\alpha_1^{+-}(T)-1}
 \quad \mbox{for}\quad\Delta > 0\,, \\
R_{<}^{+-}&=&\frac{\alpha_2^{+-}(T)-1}{\alpha_1^{zz}(T)-1}
\quad \mbox{for}\quad\Delta < 0 \,,
\end{eqnarray}
respectively. For the discussion it is useful to perform the 
unitary transformation which rotates the spins on the sublattice B around
the $z$-axis by the angle $\pi$, $\tilde{\bf S}_i={\cal U}^+{\bf S}_i{\cal U}$ with
${\cal U}=\prod_{l\in B}2 S_l^z$. We get  
$\tilde{S}_i^{x,y}=\mbox{e}^{i{\bf Q}{\bf r}_i }S_i^{x,y}$, 
$\tilde{S}_i^{z}=S_i^{z}$ and 
\begin{equation}
\label{utrafoh}
\tilde{\cal H}=\frac{1}{2}\sum_{\langle i,j\rangle}(-S_i^+S_j^- +\Delta
S_i^zS_j^z)\,.
\end{equation}
Due to $\langle A\rangle_{\cal H}=\langle \tilde{A}\rangle_{\tilde{\cal H}}$
for any operator $A$, we obtain the relations
\begin{eqnarray}
\label{utrafox1}
\chi^{+-}_{{\bf q},{\cal H}}(\omega)&=&
\chi_{{\bf k},{\tilde{\cal H}}}^{+-}(\omega)\,;\;\;
{\bf k}={\bf q}-{\bf Q}\,,\\
C^{+-}_{{\bf r},{\cal H}}&=&
\mbox{e}^{i{\bf Q r}}C_{{\bf r},{\tilde{\cal H}}}^{+-}\,,
\label{utrafox2}
\end{eqnarray}
$\chi^{zz}_{{\bf q},{\cal H}}(\omega)=
\chi_{{\bf q},{\tilde{\cal H}}}^{zz}(\omega)$, and 
$C^{zz}_{{\bf r},{\cal H}}=
C_{{\bf r},{\tilde{\cal H}}}^{zz}$. As shown
in Ref.~\cite{SFBI99}, the rotational symmetry at
$\Delta=\pm 1$ is preserved by our theory.
\bibliography{ref}
\bibliographystyle{phys}
\end{document}